\documentclass[10pt,twocolumn,letterpaper]{article}

\usepackage{cvpr}
\usepackage{times}
\usepackage{epsfig}
\usepackage{graphicx}
\usepackage{subfigure}
\usepackage{amsmath}
\usepackage{amssymb}
\usepackage{url}
\usepackage{array}
\usepackage{amsthm}
\usepackage{algorithm}               
\usepackage{algorithmic}             
\usepackage{multirow}                
\usepackage{bm}
\usepackage{footmisc}

\def\1{\mathbf{1}}
\def\0{\mathbf{0}}

\def\mA{\mathcal{A}}
\def\mB{\mathcal{B}}
\def\mM{\mathcal{M}}

\def\mC{\mathcal{C}}

\def\mO{\mathcal{O}}

\def\mE{\mathcal{E}}
\def\mR{\mathcal{R}}

\usepackage[pagebackref=true,breaklinks=true,letterpaper=true,colorlinks,bookmarks=false]{hyperref}

\cvprfinalcopy 

\def\cvprPaperID{1514} 

\ifcvprfinal\fi
\begin{document}

\title{A Model-driven Deep Neural Network for Single Image Rain Removal}
\author{Hong Wang$^{1,}$\footnotemark[1], Qi Xie$^{1,}$\footnotemark[1], Qian Zhao$^{1}$, Deyu Meng$^{2,1,}$\footnotemark[2]\\
$^{1}$Xi'an Jiaotong University; $^{2}$Macau University of Science and Technology\\
{\tt\small \{hongwang01,xq.liwu\}@stu.xjtu.edu.cn}\quad
{\tt\small timmy.zhaoqian@gmail.com}\quad
{\tt\small dymeng@mail.xjtu.edu.cn}
}
\maketitle
\renewcommand{\thefootnote}{\fnsymbol{footnote}}
\footnotetext[2]{Corresponding author}
\renewcommand{\thefootnote}{\arabic{footnote}}
\renewcommand{\thefootnote}{\fnsymbol{footnote}}
\footnotetext[1]{Equal contribution}
\renewcommand{\thefootnote}{\arabic{footnote}}
\thispagestyle{empty}

\begin{abstract}
Deep learning (DL) methods have achieved state-of-the-art performance in the task of single image rain removal. Most of current DL architectures, however, are still lack of sufficient interpretability and not fully integrated with physical structures inside general rain streaks. To this issue, in this paper, we propose a model-driven deep neural network for the task, with fully interpretable network structures. Specifically, based on the convolutional dictionary learning mechanism for representing rain, we propose a novel single image deraining model and utilize the proximal gradient descent technique to design an iterative algorithm only containing simple operators for solving the model. Such a simple implementation scheme facilitates us to unfold it into a new deep network architecture, called rain convolutional dictionary network (RCDNet), with almost every network module one-to-one corresponding to each operation involved in the algorithm. By end-to-end training the proposed RCDNet, all the rain kernels and proximal operators can be automatically extracted, faithfully characterizing the features of both rain and clean background layers, and thus naturally lead to its better deraining performance, especially in real scenarios. Comprehensive experiments substantiate the superiority of the proposed network, especially its well generality to diverse testing scenarios and good
interpretability for all its modules, as compared with state-of-the-arts both visually and quantitatively. The source codes are available at \url{https://github.com/hongwang01/RCDNet}.
\end{abstract}

\begin{figure}[t]
  \begin{center}
  \vspace{-1.5mm}
     \includegraphics[width=0.95\linewidth]{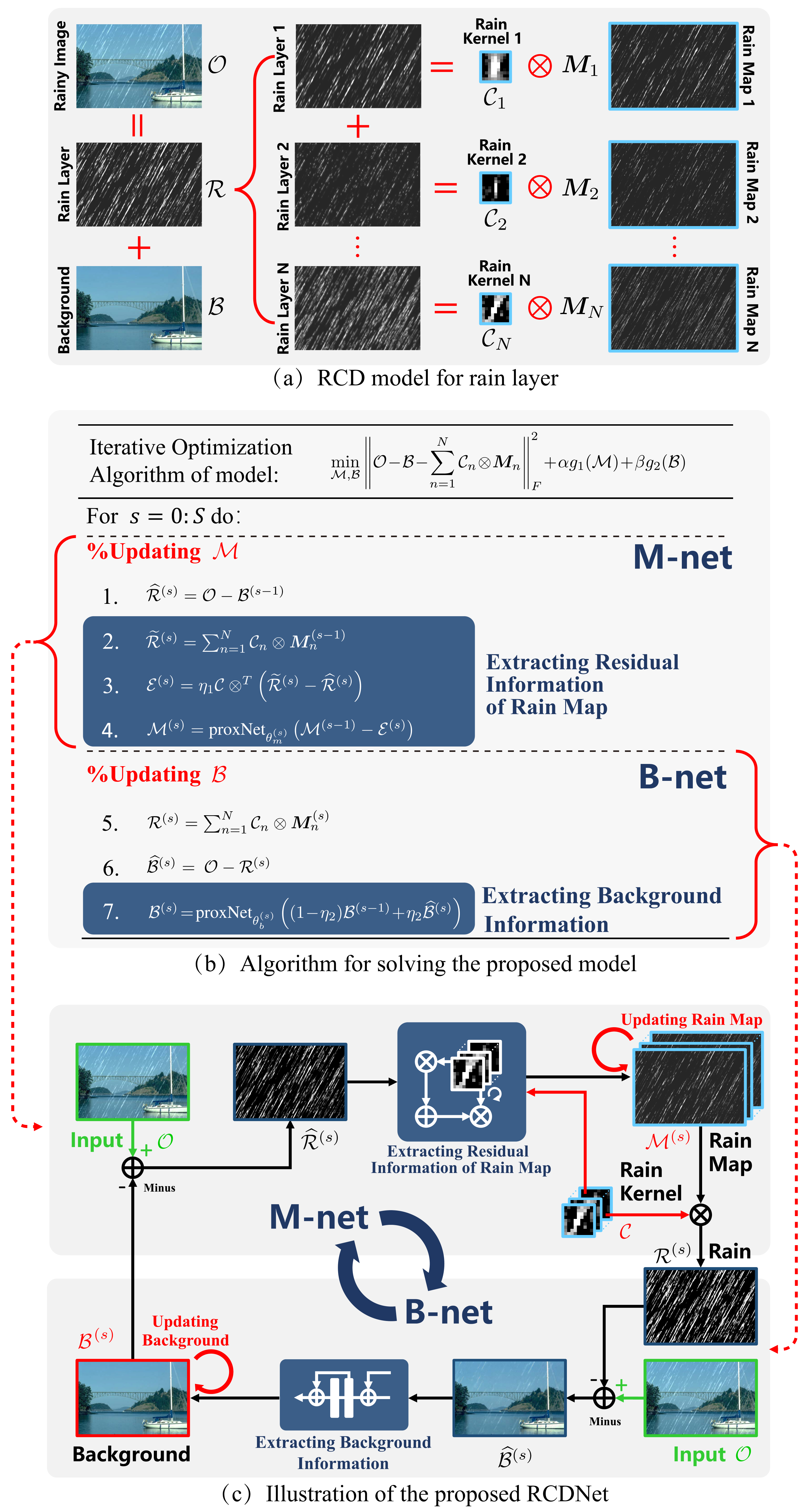}
  \end{center}
  \vspace{-3mm}
     \caption{  (a) Rain convolutional dictionary (RCD) model for rain layer.
     (b) The formulated optimization model and the corresponding iterative solution algorithm.
     (c) Visual illustration of the proposed RCDNet one-to-one corresponding to the algorithm (b).}
  \label{figintro}
    \vspace{-9mm}
  \end{figure}
\vspace{-4mm}
\section{Introduction}
Images taken under various rain conditions often suffer from unfavorable visibility, and always severely affect the performance of outdoor computer vision tasks, such as objection tracking~\cite{Comaniciu2003Kernel}, video surveillance~\cite{Shehata2008Video}, and pedestrian detection~\cite{Ludwig2009Trainable}. Hence, removing rain streaks from rainy images is an important pre-processing task and has drawn much research attention in the recent years~\cite{wang2019a,benchmark}.

In the past years, various methods have been proposed for single image rain removal task. Many researchers made focus on exploring physical properties of rain layer and background layer, and introduced various prior structures to regularize and separate them. Along this research line, the representative methods include layer priors with Gaussian mixture model (GMM)~\cite{Li2016Rain}, discriminative sparse coding (DSC)~\cite{Yu2015Removing}, and joint convolutional analysis and synthesis sparse representation (JCAS)~\cite{Gu2017Joint}. Especially, inspired by the fact that rain streaks repeatedly appear at different locations over a rainy image with similar local patterns like shape, thickness, and direction, very recently researchers represented this configuration of rain layer by the convolutional dictionary learning model~\cite{He2017Convolutional,Huang2015Convolutional}. Such a representation finely delivers this prior knowledge by imposing rain kernels (conveying repetitive local patterns) on sparse rain maps, as intuitively depicted in Fig.~\ref{figintro}~(a). These methods thus achieved state-of-the-art (SOTA) performance when the background can also be well represented, e.g., by low-rank prior in surveillance video sequences \cite{li2018video}.

Albeit effective in certain applications, the rationality of these techniques depends on the subjective prior assumptions imposed on the unknown background and rain layers to be recovered. In real scenarios, however, such learning regimes could not always adapt to different rainy images with complex, diverse, and variant structures collected from different resources. Besides, these methods generally need time-consuming iterative computations, often with efficiency issue in real applications.

Driven by the significant success of deep learning (DL) in low level vision, recent years have also witnessed the rapid progress of deep convolutional neural networks (CNN) for single image rain removal~\cite{Fu2017Clearing,zhang2018density,zhang2019image,wang2020single}. The current DL-based derainers mainly focus on designing network modules, and then train network parameters based on abundant rainy/clean image pairs to extract the background layer. Typical deraining network structures include deep detail network (DDN)~\cite{Fu2017Removing}, recurrent squeeze-and-excitation context aggregation module (RESCAN)~\cite{li2018recurrent}, progressive image deraining network (PReNet)~\cite{ren2019progressive}, spatial attentive unit (SPANet)~\cite{wang2019spatial}, and many others.

These DL strategies, however, also possess evident deficiencies. The most significant one is their weak interpretability. Network structures are always complicated and diverse, making it difficult to analyze the role of different modules and understand the underlying insights of their mechanism. Besides, most of them treat CNN as an encapsulated end-to-end mapping module without deepening into the rationality, and neglect the intrinsic prior knowledge of rain streaks such as sparsity and nonlocal similarity. This makes this methodology easily trapped into the overfitting-to-training-sample issue.

To alleviate the aforementioned issues, this paper designs an interpretable deep network, which sufficiently considers the characteristics of rain streaks and attempts to combine the advantages of the conventional model-driven prior-based and current data-driven DL-based methodologies. Specifically, our contributions are mainly three-fold:

Firstly, we propose a concise rain convolutional dictionary (RCD) model for single image by exploiting the intrinsic convolutional dictionary learning mechanism to encode rain shapes, and specifically adopt the  proximal gradient technique~\cite{beck2009fast} to  design an optimization algorithm for solving it. Different from traditional solvers for the RCD model containing complex operations (e.g., Fourier transformation), the algorithm only contains simple computations (see Fig.~\ref{figintro}~(b)) easy to be implemented by general network modules. This facilitates our algorithm capable of being easily unfolded into a deep network architecture.

Secondly, by unfolding the algorithm, we design a new deep network architecture for image deraining, called RCDNet. The specificity of this network lies on its exact step-by-step corresponding relationship between its modules and the algorithm operators, and thus successively possesses the interpretability of all its modules as that of all steps in the algorithm. Specifically, as shown in Fig.~\ref{figintro}~(b) and (c), each iteration of the algorithm contains two sub-steps, respectively updating the rain map (convoluted by the learned rain kernels) and background layer, and each stage of the RCDNet also contains two sub-networks (M-net and B-net). Each output of the intermediate layer in the network is thus with clear interpretation, which greatly facilitates a deeper analysis on what happens inside the network during training, and a comprehensive understanding why the network works or not (as the analysis presented in Sec. 5.2).

Thirdly, comprehensive experimental results substantiate the superiority of the RCDNet beyond SOTA conventional prior-based and current DL-based methods both quantitatively and visually. Especially, attributed to its well interpretability, not only the underlying rationality and insights of the network can be intuitively understood through visualizing the amelioration process (like the gradually rectified background and rain maps) over all network layers by general users, but also the network can yield generally useful rain kernels for expressing rain shapes and proximal operators for delivering the prior knowledge of background and rain maps for a rainy image, facilitating their general availability to more real-world rainy images.

The paper is organized as follows. Sec. 2 reviews the related rain removal work. Sec. 3 presents the RCD model for rain removal as well as the algorithm designed for solving it. Then Sec. 4 introduces the unfolding deep network for the algorithm. The experimental results are demonstrated in Section 5 and the paper is finally concluded.

\section{Related work}
In this section, we give a brief review on the most related work on rain removal for images. Depending on the input data, the existing algorithms can be categorized into two groups: video based and single image based ones.

\subsection{Video deraining methods}
Garg and Nayar \cite{Garg2004Detection} first tried to analyze the visual effects of raindrops on imaging systems, and utilized a space-time correlation model to capture the dynamics of raindrops and a physics-based motion blur model to illustrate the photometry of rain. For better visual quality, they further proposed to increase the exposure time or reduce the depth of field of a camera~\cite{garg2007vision,garg2005does}. Later, both temporal and chromatic properties of rain were considered and then background layer was extracted from rainy video by utilizing different strategies such as K-means clustering~\cite{zhang2006rain}, Kalman filter~\cite{park2008rain}, and GMM~\cite{bossu2011rain}. Besides, a spatio-temporal frequency based raindrop detection method was provided in~\cite{barnum2010analysis}.

In recent years, researchers introduced more intrinsic characteristics of rainy video to the task, e.g., similarity and repeatability of rain streaks~\cite{Chen2013A}, low-rankness among multi-frames~\cite{Jin2015Video}, and sparsity and smoothness of rain streaks~\cite{Jiang2017A}. To handle heavy rain and dynamic scenes, a matrix decomposition based video deraining algorithm was presented in~\cite{Ren2017Video}. Afterwards, rain streaks were encoded as a patch based GMM to adapt a wider range of rain variations~\cite{wei2017should}. More characteristics of rain streaks in a rainy video were explored including repetitive local patterns and multi-scale configurations and they were described as multiscale convolutional sparse coding model~\cite{li2018video}. More recently, there are some DL-based methods proposed for this task. Chen~\emph{et al.}~\cite{Jie2018Robust} presented a CNN architecture and utilized superpixel to handle torrential rain fall with opaque streak occlusions. To further improve visual quality, Liu~\emph{et al.}~\cite{liu2018erase} designed a joint recurrent rain removal and reconstruction network that integrated rain degradation classification, rain removal, and background details reconstruction. To handle dynamic video contexts, they further developed a dynamic routing residue recurrent network~\cite{Liu2018D3R}. Though these methods work well for videos, they cannot directly perform in single image cases due to the lack of temporal knowledge.

\subsection{Single image deraining methods}
Compared with video deraining task under a sequence of images, rain removal from a single image is much more challenging. The early attempts utilized the model-driven strategies by decomposing a single rainy image into low frequency part (LFP) and high frequency part (HFP) and then specifically extracted rain layer from the HFP based on various processing such as guided filter~\cite{Ding2016Single,Jing2012Removing} and nonlocal means filtering~\cite{Kim2014Single}. Later, researchers made more focus on exploring the prior knowledge of rain and rain-free layers of a rainy image, and designing proper regularizer to extract and separate them~\cite{Kang2012Automatic,sun, Yu2015Removing,Li2016Rain, Wang2017A, Zhu2017Joint}. E.g., \cite{Gu2017Joint} considered the specific sparsity characteristics of rain-free and rain parts and expressed them as the joint analysis and synthesis sparse representation models, respectively. \cite{He2017Convolutional} used a similar manner to deliver local repetitive patterns of rain streaks across the image as the RCD model. Albeit achieving good performance on certain scenarios, these prior-based methods rely on the subjective prior assumptions, while could not always generally work well for practical complicated and highly diverse rain shapes in real rainy images collected from different resources.

Recently, a number of DL-based single image rain streak removal methods were proposed through constructing diverse network modules~\cite{Fu2017Clearing,Fu2017Removing,li2018recurrent,zhang2018density,zhang2019image}. To handle heavy rain, Yang~\emph{et al.}~\cite{Yang2019Joint} developed a multi-stage joint rain detection and estimation network for single image (JORDER\_E). Very recently, Ren~\emph{et al.}~\cite{ren2019progressive} designed a PReNet that repeatedly unfolded several Resblocks and a LSTM layer. Wang~\emph{et al.}~\cite{wang2019spatial} presented an attention unit based SPANet for removing rain in a local-to-global manner. Through using abundant rainy/clean image pairs to train the deep model, these methods achieve favorable visual quality and SOTA quantitative measures of derained results. Most of these methods, however, just utilize network modules assembled with some off-the-shelf components in current DL toolkits to directly learn background layer in an end-to-end way, and largely ignore the intrinsic prior structures inside the rain streaks. This makes them lack of evident interpretability in their network architectures and still have room for further performance enhancement.

At present, there is a new type of single image derainers that try to combine prior and DL methodologies. For example, Mu~\emph{et al.}~\cite{Mu2019Learning} utilized CNN to implicitly learn prior knowledge for background and rain streaks, and formulated them into traditional bi-layer optimization iterations. Wei~\emph{et al.}~\cite{wei2019semi} provided a semi-supervised rain removal method (SIRR) that described rain layer prior as a general GMM and jointly trained the backbone--DDN. Albeit obtaining initial success, they still use CNN architectures as their main modules to construct the network, which is thus still lack of sufficient interpretability.

\section{RCD model for single image deraining}
\subsection{Model formulation}
For a observed color rainy image denoted as $\mO\in\mathbb{R}^{H\times W\times 3}$, where $H$ and $W$ are the height and width of the image, respectively, it can be rationally separated as:
\begin{equation}\label{e1}
\mO=\mB+\mR,
\end{equation}
where $\mB$ and $\mR$ represent the background and rain layers of the image, respectively. Then, the aim of most of DL-based deraining methods is to estimate the mapping function (expressed by a deep network) from $\mO$ to $\mB$ (or $\mR$).

Instead of heuristically constructing a complex deep network architecture, we first consider the problem under the conventional prior-based methodology through exploiting the prior knowledge for representing rain streaks \cite{Gu2017Joint, He2017Convolutional,li2018video}. Specifically, as shown in Fig.~\ref{figintro}~(a), by adopting the RCD mechanism, the rain layer can be modeled as:
\begin{equation}\label{theR}
  \mR^c =  \sum_{n=1}^{N} \bm{C}^c_{n}  \otimes \bm{M}_{n}, c=1,2,3,
\end{equation}
where $\mR_c$ denotes the $c^{\text{th}}$ color channel of $\mR$, and $\left\{\bm{C}^c_{n}\right\}_{n,c} \subset \mathbb R^{k \times k}$ is a set of rain kernels which describes the repetitive local patterns of rain streaks, and $\left\{\bm{M}_{n}\right\}_{n} \subset \mathbb R^{H \times W}$ represents the corresponding rain maps representing the locations where local patterns repeatedly appear. $N$ is the number of kernels and $\otimes$ is the 2-dimensional (2D) convolutional operation. For conciseness, we rewrite (\ref{theR}) as $  \mR = \sum^{N}_{n=1} \mC_{n}\otimes \bm{M}_{n}$ throughout the paper, where $\mathcal{C}_{n}\in\mathbb{R}^{k\times k\times 3}$ is the tensor form of $\bm{C}^c_n$s and the convolution is performed  between $\mC_n$ and the matrix $\bm{M}_n$ one channel by one channel.
Then,  we  can rewrite the model (\ref{e1}) as:
\begin{equation}\label{e2}
{\mO}=\mB+ \sum_{n=1}^{N} \mC_{n}  \otimes \bm{M}_{n}.
\end{equation}

It should be noted that the rain kernels actually can be viewed a set of convolutional dictionary \cite{Huang2015Convolutional} for representing repetitive and similar local patterns underlying rain streaks, and a small number of rain kernels can finely represent wide range of rain shapes\footnote{We simply set $N=32$ for all our experiments.}. They are common knowledge for representing different rain types across all rainy images, and thus could be learned from abundant training data by virtue of the strong learning capability of end-to-end training manner of deep learning (see more details in Sec. 4).
Unlike rain kernels, the rain maps must vary with the input rainy image as the locations of rain streaks are totally random.
Therefore, for predicting the clean image from a testing input rainy one, the key issue is to output $\bm{M}_n$s and $\mB$ from $\mO$ with the rain kernels $\mC_n$s fixed, and the corresponding optimization problem is:
\begin{equation}\label{o21}
\min_{{\mM,\mB}}\left\|\mO\!-\!\mB\!-\!\sum_{n=1}^{N} \mC_{n} \!\otimes\! \bm{M}_{n}\right\|_{F}^{2}\!+\!\alpha g_{1}(\mM)\!+\!\beta g_{2}(\mB),
\end{equation}
where $\mM\in\mathbb{R}^{H\times W\times N}$  is the tensor form of $\bm{M}_n$s. $\alpha$ and $\beta$ are trade-off parameters. $g_{1}(\cdot)$ and $g_{2}(\cdot)$ mean the regularizers to deliver the prior structures of $\bm{M}_n$ and $\mB$, respectively.

  \begin{figure*}[t]
  \begin{center}
  \vspace{-2mm}
     \includegraphics[width=1\linewidth]{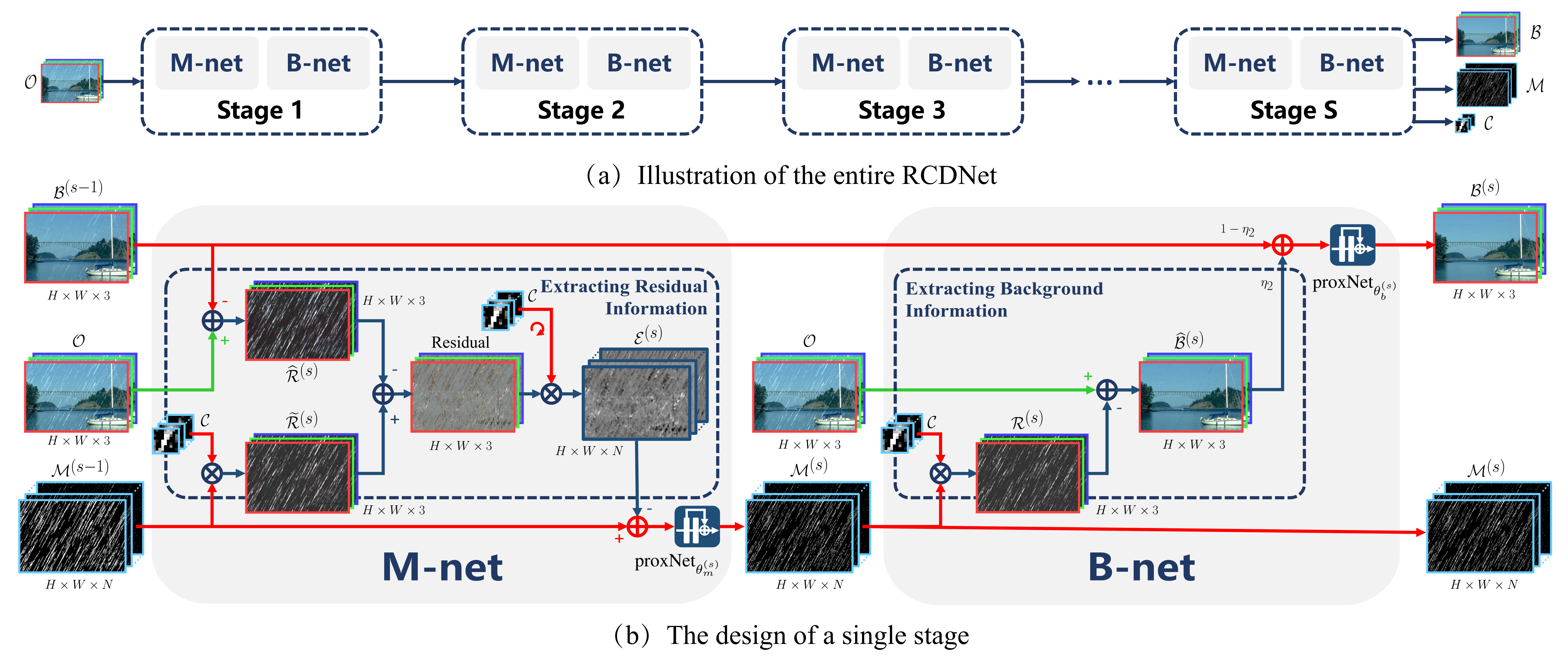}
  \end{center}
  \vspace{-3mm}
     \caption{(a) The proposed network with $S$ stages. The network takes a rainy image $\mathcal{O}$ as input and outputs the learned rain kernel $\mathcal{C}$, rain map $\mathcal{M}$, and clean background image $\mathcal{B}$. (b) Illustration of the network architecture at the $s^{\text{th}}$ stage. Each stage consists of M-net and B-net to accomplish the update of rain map $\mathcal{M}$ and background layer $\mathcal{B}$, respectively. The images are better to be zoomed in on screen.}
  \label{figflow}
    \vspace{-5mm}
  \end{figure*}

\subsection{Optimization algorithm}
Since we want to build a possibly perfect step-by-step corresponding deep unfolding network architecture for solving the problem~(\ref{o21}), it is critical to build an algorithm which contains only simple computations easy to be transformed to network modules. The traditional solvers for RCD-based model usually contain certain complicated operations, e.g., the Fourier transform and inverse Fourier transform~\cite{Huang2015Convolutional, WohlbergEfficient,li2018video}, which are hard to accomplish such exact transformation from algorithm to network structure. We thus prefer to build a new algorithm for solving the problem through alternately updating $\mM$ and $\mB$ by proximal gradient method~\cite{beck2009fast}. In this manner, only simple computations can be involved. The details are as follows:

\textbf{Updating $\mM$}: The rain maps $\mM$ can be updated by solving the quadratic approximation~\cite{beck2009fast} of the problem (\ref{o21}) as:
\small
\begin{equation}\label{minq1}
  \!\min_{\mM}\! \frac{1}{2} \!\left\| \mM \!-\!\! \left(\! \mM^{(s-1)}\!\!-\! \eta_{1}\nabla f\left(\mM^{(s-1)}\right)  \!\right) \right\|_F^2 \!+ \alpha\eta_{1} g_{1}\left( \mM \right),
\end{equation}
\normalsize
where $\mM^{(s-1)}$ is the updating result of the last iteration, $\eta_{1}$ is the stepsize parameter, and  $f\left(\mM^{(s-1)}\right)\!=\! \left\|\mO\!-\!\mB^{(s-1)}\!-\!\sum_{n=1}^{N} {\mC}_{n} \!\otimes\! \bm{M}^{(s-1)}_n\right\|_{F}^{2}$. Corresponding to general regularization terms \cite{donoho1995noising}, the solution of  Eq. (\ref{minq1}) is:
\begin{equation}\label{solm}
 \mM^{(s)} = \mbox{prox}_{\alpha\eta_{1}}\left(\! \mM^{(s-1)} \!-\! \eta_{1}\nabla f\left(\mM^{(s-1)}\right)  \!\right).
\end{equation}
Moreover, by  substituting
\small
\begin{equation}\label{gradf}
  \nabla\! f\!\left(\!\mM^{(s-1)}\!\right)\!=\!\mathcal{C}\!\otimes^{T}\!\left(\sum_{n=1}^{N}\mathcal{C}_{n} \!\otimes\! \bm{M}_{n}^{(s-1)}\!+\!\mB^{(s-1)}\!-\!\mO\!\right),
\end{equation}
\normalsize
where $\mC\in\mathbb{R}^{k\times k\times N \times 3}$ is a 4-D tensor stacked by $\mC_n$s, and $\otimes^T$ denotes the transposed convolution\footnote{For any tensor $\mA\in\mathbb{R}^{H\times W \times 3}$, we can calculate the $n^{\text{th}}$ channel of $\mathcal{C}\!\otimes^{T}\!\mA$ by $\sum_{c = 1}^3 \mC_{\{:,:,n,c\}} \otimes^{T} \mA_{\{:,:,c\}}$. },
we can obtain the updating formula for $\mM$ as\footnote{It can be proved that, with  small enough $\eta_1$ and $\eta_2$,  Eq. (\ref{updatem1}) and Eq. (\ref{updateb1}) can both lead to the reduction of objective function (\ref{o21})~\cite{beck2009fast}. \label{fn:repeat1}}:
\small
\begin{equation}\label{updatem1}
\begin{split}
  &\mM^{(s)}= \!\!\\&\!\!
  \mbox{prox}_{\alpha\eta_{1}}\!\!\left(\!\! \mM^{(s-1)}
  \!\!-\!\! \eta_{1}\mathcal{C}\!\otimes^{T}\!\!\!\left(\sum_{n=1}^{N} \!\mathcal{C}_{n} \!\!\otimes\! \bm{M}_{n}^{(s-1)}\!\!\!+\!\mB^{(s-1)}\!\!\!-\!\mO\!\right) \!\right)\!,
\end{split}
\end{equation}
\normalsize
where $\mbox{prox}_{\alpha\eta_1}(\cdot)$ is the proximal operator dependent on the regularization term $g_{1}(\cdot)$ with respect to $\mM$. Instead of choosing a fixed regularizer in the model, the form of the proximal operator can be automatically learned from training data. More details will be presented in the next section.

\textbf{Updating $\mB$}:
Similarly, the quadratic approximation of the problem (\ref{o21}) with respect to $\mB$  is:
\begin{equation}\label{subProblem2b}
  \!\!\min_{\mB}\!\frac{1}{2}\!\left\| \mB \!-\! \left(\! \mB^{(s-1)} \!-\! \eta_{2}\nabla h\left(\mB^{(s-1)}\right)  \!\right) \right\|_F^2 \!\!+\! \beta\eta_{2} g_{2}\! \left( \mB \right).
\end{equation}
where
$
  \nabla h\left(\!\mB^{(s-1)}\!\right) = \sum_{n=1}^{N} \mC_{n} \otimes \bm{M}_{n}^{(s)}+\mB^{(s-1)}-\mO,
$
and it is easy to deduce that  the final updating rule for $\mB$ is\footref{fn:repeat1}:
\small
\begin{equation}\label{updateb1}
\begin{split}
  &\mB^{(s)} \!=\!\\
  &\mbox{prox}_{\beta\eta_{2}}\!\!\left(\!\left(1-\eta_2\right) \mB^{(s-1)}
  \!\!+\!\eta_{2}\!\!\left(\!\mO\! - \!\sum_{n=1}^{N} \mC_{n} \!\otimes\! \bm{M}_{n}^{(s)}\!\right)\!\right)\!.
\end{split}
\end{equation}
\normalsize
where $\mbox{prox}_{\beta\eta_2}(\cdot)$ is the  proximal operator correlated to the regularization term $g_{2}(\cdot)$ with respect to $\mB$.

Based on this iterative algorithm, we can then construct our deep unfolding network as follows.

\section{The rain convolutional dictionary network}
Inspired by the recently raised deep unfolding techniques in various tasks such as deconvolution~\cite{zhang132017learning}, compressed sensing~\cite{yang2017admm}, and dehazing~\cite{yang2018proximal}, we build a network structure for single image rain removal task by unfolding each iterative steps of the aforementioned algorithm as the corresponding network module. We especially focus on making all network modules one-to-one corresponding to the algorithm implementation operators, for better interpretability.

As shown in Fig.~\ref{figflow}~(a), the proposed network consists of $S$ stages, corresponding to $S$ iterations of the algorithm for solving ~(\ref{o21}). Each stage achieves the sequential updates of $\mM$ and $\mB$ by M-net and B-net. As displayed in Fig.~\ref{figflow}~(b), exactly corresponding to each iteration of the algorithm,
 in each stage of the network, M-net takes the observed rainy image $\mO$ and the previous outputs $\mB^{(s-1)}$ and $\mM^{(s-1)}$ as inputs, and outputs an updated $\mM^{(s)}$, and then B-net takes $\mO$ and $\mM^{(s)}$ as inputs, and outputs an updated $\mB^{(s)}$.

\subsection{Network design}
The key issue of unrolling the algorithm here is how to represent the two proximal operators involved in (\ref{updatem1}) and (\ref{updateb1}) while other operations can be naturally performed with generally used operators in normal networks~\cite{paszke2017automatic}. In this work, we simply choose a ResNet \cite{he2016deep} to construct the two proximal operators as many other works did~\cite{xie2019multispectral,yang2018proximal}.
Then, we can separately decompose the updating rules for $\mM$ as (\ref{updatem1}) and $\mB$ as (\ref{updateb1}) into sub-steps and achieve the following procedures for the $s^{\text{th}}$ stage of the RCDNet:
\small
\begin{eqnarray}\label{unfoldm}
\hspace{-7.5mm}\text{M-net}: \left\{\begin{matrix}
\hspace{-25mm}\widehat{\mR}^{(s)}=\mO-\mB^{(s-1)},\\
\hspace{-13.5mm}\widetilde{\mR}^{(s)}=\sum_{n=1}^{N}\mC_{n}\otimes\bm{M}_{n}^{(s-1)},\\
\hspace{-8.5mm}\mathcal{E}^{(s)}=\eta_{1}\mC\otimes^{T}\left(\widetilde{\mR}^{(s)}-\widehat{\mR}^{(s)}\right),\\
\mM^{(s)}=\text{proxNet}_{\theta_m^{(s)}}\left(\mM^{(s-1)}-\mathcal{E}^{(s)}\right),
\end{matrix}\right.
\end{eqnarray}
\vspace{-4mm}
\begin{eqnarray}\label{unfoldb}
\text{B-net}: \left\{\begin{matrix}
 \hspace{-25.5mm}{\mR}^{(s)}=\sum_{n=1}^{N}\mC_{n}\otimes\bm{M}_{n}^{(s)},\\
 \hspace{-36.5mm}\widehat{\mB}^{(s)}=\mO-{\mR}^{(s)},\\
\mB^{(s)}\!=\!\text{proxNet}_{\theta_b^{(s)}}\left((1\!-\!\eta_2)\mB^{(s-1)}\!+\!\eta_2\widehat{\mB}^{(s)}\right),
\end{matrix}\right.
\end{eqnarray}\normalsize
where $\text{proxNet}_{\theta_m^{(s)}}(\cdot)$ and $\text{proxNet}_{\theta_b^{(s)}}(\cdot)$  are two ResNets consisting of several Resblocks with the parameters ${\theta_m^{(s)}}$ and ${\theta_b^{(s)}}$ at the $s^{\text{th}}$ stage, respectively.

We can then design the network architecture, as shown in Fig. \ref{figflow}, by transforming the operators in (\ref{unfoldm}) and (\ref{unfoldb}) step-by-step. All the parameters involved can be automatically learned from training data in an end-to-end manner, including $\{\theta_m^{(s)},\theta_b^{(s)}\}_{s=1}^{S}$, rain kernels $\mC$, $\eta_1$, and $\eta_2$.

It should be indicated that both of the two sub-networks are very interpretable. As shown in Fig. \ref{figflow}~(b),  the M-net accomplishes the extraction of residual information $\mE^{(s)}$ of rain maps. Specifically, $\widehat{\mR}^{(s)}$ is the rain layer estimated with the previous background $\mB^{(s-1)}$, and $\widetilde{\mR}^{(s)}$ is the rain layer achieved by the generative model (\ref{theR}) with the estimated $\mM^{(s-1)}$. Then the M-net calculates the residual information between the two rain layers obtained in this two ways, and extracts the residual information $\mE^{(s)}$ of rain maps with the transposed convolution of rain kernels to update the rain map.
Next, the B-net recovers the background $\widehat{\mB}^{(s)}$ estimated with current rain kernel and rain maps $\mM^{(s)}$, and fuses this estimated $\widehat{\mB}^{(s)}$ with the previously estimated $\mB^{(s-1)}$ by weighted parameters $\eta_2$ and ($1-\eta_2$) to get the updated background ${\mB}^{(s)}$. Here, we set $\mM^{(0)}$ as 0 and initialize $\mB^{(0)}$ by a convolutional operator on $\mO$\footnote{More network design details are described in supplemental file.}.

\emph{\textbf{Remark:}} From Fig.~\ref{figflow}, the input tensor of $\text{proxNet}_{\theta_b^{(s)}}(\cdot)$ has the same size ${H\times W\times 3}$ as the to-be-estimated $\mB$. Evidently, this is not beneficial for learning $\mB$ since most of the previous updating information would be compressed due to few channels. To better keep and deliver image features, we simply expand the input tensor at the $3^{\text{rd}}$ mode for more channels in experiments (see more in supplemental file).

\subsection{Network training}
\textbf{Training loss.} For simplicity, we adopt the mean square error (MSE)~\cite{Jing2012Removing} for the learned background and rain layer at every stage as the training objective function:
\begin{equation}\label{Loss}
  L = \sum_{s=0}^{S}\lambda_{s}\left\|\mathcal{B}^{(s)}\!-\!\mathcal{B} \right\|_F^2\!+\!\sum_{s=1}^{S}\gamma_{s}\left\| {\mathcal{O}\!-\!\mathcal{B}\!-\!\mathcal{R}}^{(s)} \right\|_F^2,
\end{equation}
where $\mathcal{B}^{(s)}$ and $\mathcal{R}^{(s)}$ separately denote the derained result and extracted rain layer as expressed in~(\ref{unfoldb}) at the $s^{\text{th}}$ stage ($s=0,1,\cdots,S$). $\lambda_{s}$ and $\gamma_{s}$ are tradeoff parameters\footnote{In all experiments, we simply set $\lambda_{S}=\gamma_{S}=1$ to make the outputs at the final stage play a dominant role, and other parameters as 0.1 to help find the correct parameter in each stage. More parameter settings are discussed in supplementary material.}.

\textbf{Implement details.}
We implement our network based on a NVIDIA GeForce GTX 1080Ti GPU. We adopt the Adam optimizer~\cite{Kingma2014Adam} with the batch size of 16 and the patch size of 64$\times$64. The initial learning rate is $1\times$$10^{-3}$ and divided by 5 every 25 epochs. The total epoch is 100.

\section{Experimental results}
We first conduct ablation study and model visualization to verify the underlying mechanism of the proposed network, and then present experiments on synthesized benchmark datasets and real datasets for performance evaluation.

\subsection{Ablation study}
\textbf{Dataset and performance metrics.} In this section, we use Rain100L to perform all the ablation studies. The synthesized dataset consists of 200 rainy/clean image pairs for training and 100 pairs for testing~\cite{Yang2019Joint}. Two performance metrics are employed, including peak-signal-to-noise ratio (PSNR)~\cite{Huynh2008Scope} and structure similarity (SSIM)~\cite{wang2004image}. Note that as the human visual system is sensitive to the Y channel of a color image in YCbCr space, we compute PSNR and SSIM based on this luminance channel.
\begin{table}[t]
    \vspace{-1mm}
\centering
\caption{Effect of stage number $S$ on the performance of RCDNet.}
\footnotesize
\setlength{\tabcolsep}{1.3pt}
\begin{tabular}{c|c|c|c|c|c|c|c|c}
\hline
Stage No. & $S$=0 & $S$=2 & $S$=5 & $S$=8 & $S$=11 & $S$=14& $S$=17 & $S$=20\\
\hline
PSNR & 35.93 &38.46 &39.35  &39.60 &39.81 &39.90 &40.00 &39.91\\
\hline
SSIM  &0.9689 &0.9813 &0.9842 &0.9850 &0.9855 &0.9858 &0.9860 & 0.9858\\
\hline
\end{tabular}
\label{tabS1}
    \vspace{-2mm}
\end{table}

Table~\ref{tabS1} reports the effect of stage number $S$ on deraining performance of our network. Here, $S=0$ means that the initialization $\mathcal{B}^{(0)}$ is directly regraded as the recovery result. Taking $S=0$ as a baseline, it is seen that only with 2 stages, our method achieves significant rain removal performance, which validates the essential role of the proposed M-net and B-net. We also observe that when $S=20$, its deraining performance is slightly lower than that of $S=17$ since larger $S$ would make gradient propagation more difficult. Based on such observation, we easily set $S$ as 17 throughout all our experiments. More ablation results and discussions are listed in supplementary material.
\begin{figure}[t]
  \begin{center}
     \includegraphics[width=1\linewidth]{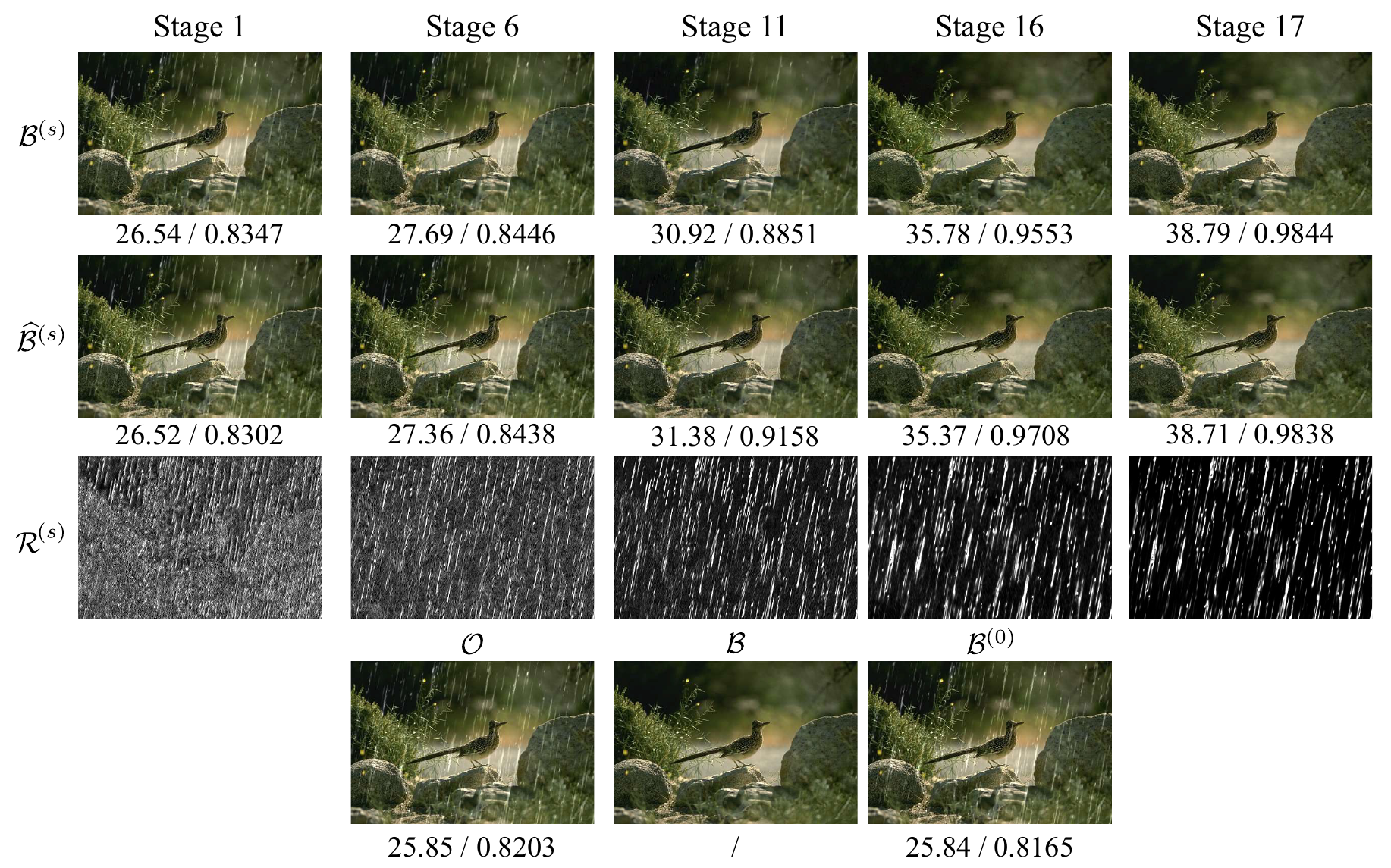}
  \end{center}
  \vspace{-3mm}
     \caption{Visualization of the recovery background $\mB^{(s)}$, $\widehat{\mB}^{(s)}$ as expressed in Eq.~(\ref{unfoldb}), and the rain layer ${\mR}^{(s)}$  at different stages. The stage number $S$ is 17. PSNR/SSIM for reference. The images are better to be zoomed in on screen.}
  \label{figverstage}
    \vspace{-1mm}
\end{figure}

\begin{figure}[t]
  \begin{center}
  \vspace{-2mm}
     \includegraphics[width=1\linewidth]{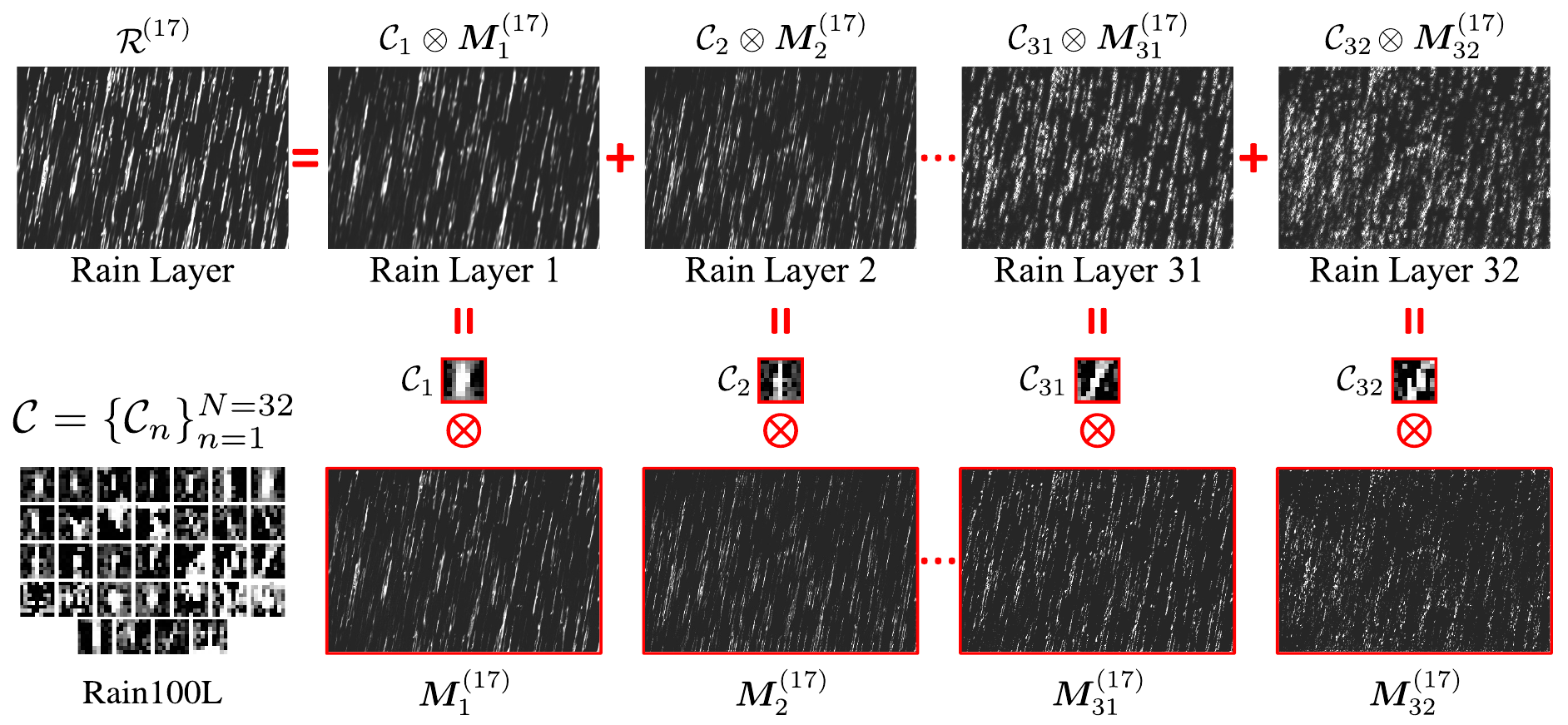}
  \end{center}
  \vspace{-3mm}
     \caption{At the final stage $s=17$, the extracted rain layer, rain kernels ${\mC_{n}}$, and rain maps $\bm{M_{n}}$ for the input $\mO$ in Fig.~\ref{figverstage}. The lower left is the rain kernels ${\mC}$ learned from Rain100L. The images are better to be zoomed in on screen.}
  \label{figvercm}
    \vspace{-5mm}
\end{figure}
\subsection{Model verification}
We then show how the interpretability of this RCDNet facilitates an easy analysis for the working mechanism inside the network modules.
\begin{figure*}[t]
  \begin{center}
  \vspace{-2mm}
     \includegraphics[width=0.95\linewidth]{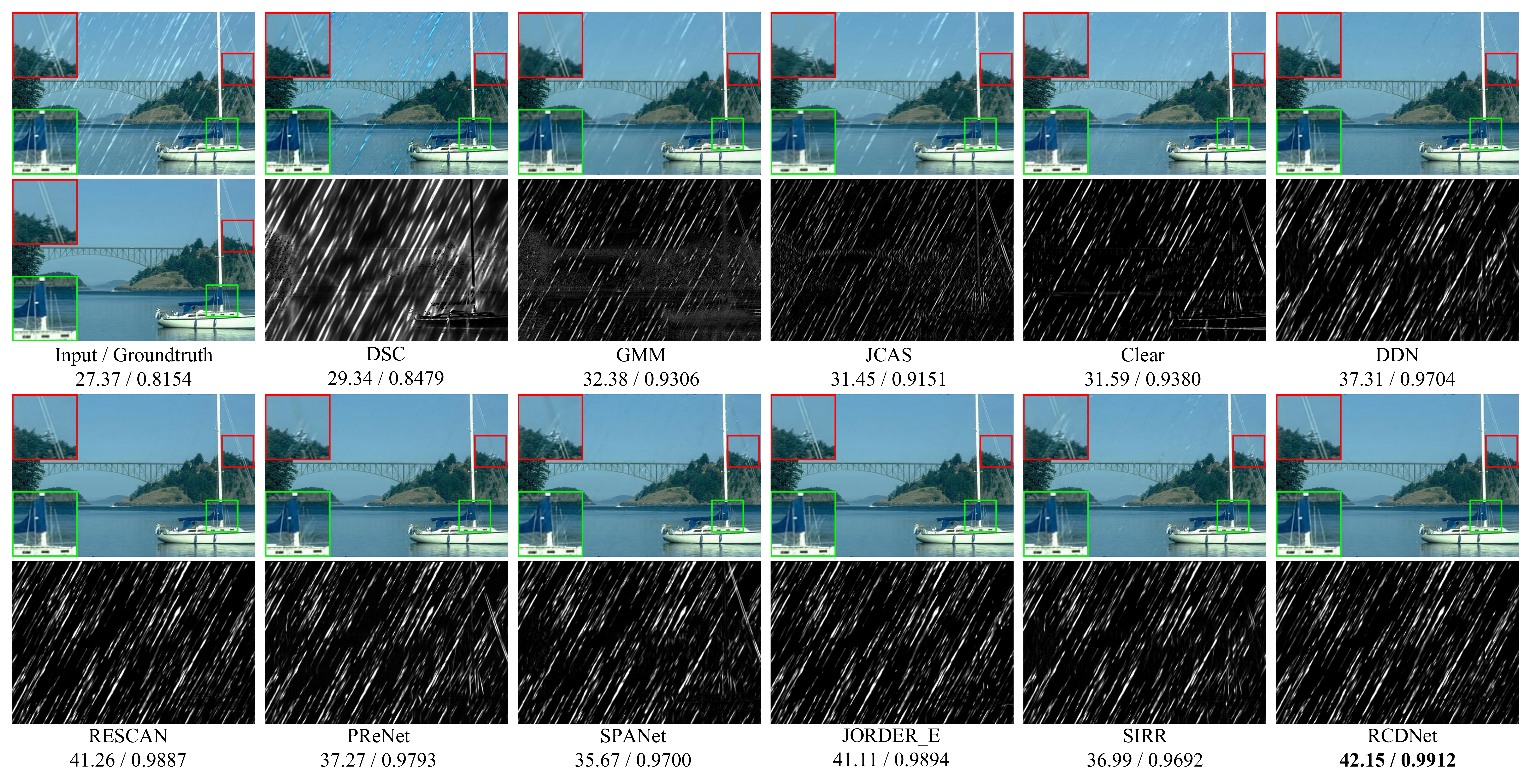}
  \end{center}
  \vspace{-5mm}
     \caption{$1^{\text{st}}$ column: input rainy image (upper) and groundtruth (lower). $2^{\text{nd}}$-$12^{\text{th}}$ column: derained results (upper) and extracted rain layers (lower) by 11 competing methods. PSNR/SSIM for reference. Bold indicates top $1^{\text{st}}$ rank.}
  \label{figrainl}
    \vspace{-5mm}
\end{figure*}

Fig.~\ref{figverstage} presents the extracted background layer $\mB^{(s)}$ ($1^{\text{st}}$ row), $\widehat{\mB}^{(s)}$($2^{\text{nd}}$ row) that represents the role of M-net in helping restore clean background, and rain layer ${\mR}^{(s)}$ ($3^{\text{rd}}$ row) at different stages. We can find that with the increase of $s$, ${\mR}^{(s)}$ covers more rain streaks and fewer image details, and $\widehat{\mB}^{(s)}$ and $\mB^{(s)}$ are also gradually ameliorated. These should be attributed to the proper guidance of the RCD prior for rain streaks and the mutual promotion of M-net and B-net that enables the RCDNet to be evolved to a right direction.

Fig.~\ref{figvercm} presents the learned rain kernels and the rain maps for the input $\mO$ in Fig.~\ref{figverstage}. Clearly, the RCDNet finely extracts proper rain layers explicitly based on the RCD model. This not only verifies the reasonability of our method but also manifests the peculiarity of our proposal. On one hand, we utilize a M-net to learn sparse rain maps instead of directly learning rain streaks that makes learning process easier. On the other hand, we exploit training data to automatically learn rain kernels representing general repetitive local patterns of rain with diverse shapes. This facilitates
their general availability to more real-world rainy images.
\begin{table}[h]
\vspace{-1mm}
\centering
\caption{PSNR and SSIM comparisons on four benchmark datasets. Bold and bold italic indicate top $1^{\text{st}}$ and $2^{\text{nd}}$ rank, respectively.}
\footnotesize
\setlength{\tabcolsep}{0.8pt}
\begin{tabular}{c|cc|cc|cc|cc}
\hline
  Datasets & \multicolumn{2}{|c|}{Rain100L} & \multicolumn{2}{|c@{}}{Rain100H} & \multicolumn{2}{|c|}{Rain1400} & \multicolumn{2}{|c@{}}{Rain12}\\
\hline
  Metrics & PSNR & SSIM & PSNR & SSIM  & PSNR & SSIM & PSNR & SSIM\\
\hline
  Input & 26.90 & 0.8384 & 13.56 & 0.3709 & 25.24 & 0.8097 & 30.14 & 0.8555\\
\hline
  DSC\cite{Yu2015Removing} & 27.34 & 0.8494 & 13.77 & 0.3199  & 27.88 &0.8394 & 30.07 &0.8664\\
\hline
  GMM\cite{Li2016Rain} &29.05 &0.8717 & 15.23 &0.4498  &27.78 & 0.8585 & 32.14 & 0.9145 \\
\hline
  JCAS\cite{Gu2017Joint}  & 28.54 & 0.8524 & 14.62 & 0.4510 &26.20 & 0.8471 & 33.10 &0.9305 \\
\hline
  Clear\cite{Fu2017Clearing} &30.24 & 0.9344 & 15.33 & 0.7421 & 26.21& 0.8951 & 31.24 & 0.9353\\
\hline
  DDN\cite{Fu2017Removing}& 32.38 & 0.9258 & 22.85 & 0.7250 & 28.45 & 0.8888 & 34.04 & 0.9330 \\
\hline
  RESCAN\cite{li2018recurrent}   & 38.52& 0.9812 &29.62 & 0.8720 &32.03& 0.9314 &36.43&0.9519\\
\hline
  PReNet\cite{ren2019progressive}& 37.45& 0.9790 &30.11& \textit{\textbf{0.9053}} & \textit{\textbf{32.55}} & \emph{\textbf{0.9459}}& 36.66& 0.9610\\
\hline
  SPANet\cite{wang2019spatial} & 35.33 & 0.9694 &25.11 & 0.8332 & 29.85& 0.9148 & 35.85& 0.9572 \\
\hline
  JORDER\_E\cite{Yang2019Joint} &\emph{\textbf{38.59 }}&\textit{\textbf{0.9834}}& \textit{\textbf{30.50}} &0.8967 &32.00 & 0.9347 &\emph{\textbf{36.69 }}&\emph{\textbf{0.9621}}\\
\hline
  SIRR\cite{wei2019semi} & 32.37 & 0.9258 & 22.47 & 0.7164 & 28.44 & 0.8893 & 34.02& 0.9347\\
\hline
RCDNet & \textbf{40.00} & \textbf{0.9860} & \textbf{31.28} & \textbf{0.9093} & \textbf{33.04} & \textbf{0.9472} & {\textbf{37.71}} &{\textbf{0.9649}}\\
\hline
\end{tabular}
\label{tabsyn1}
\vspace{-5mm}
\end{table}
\begin{figure*}[t]
  \begin{center}
  \vspace{-2mm}
     \includegraphics[width=0.82\linewidth]{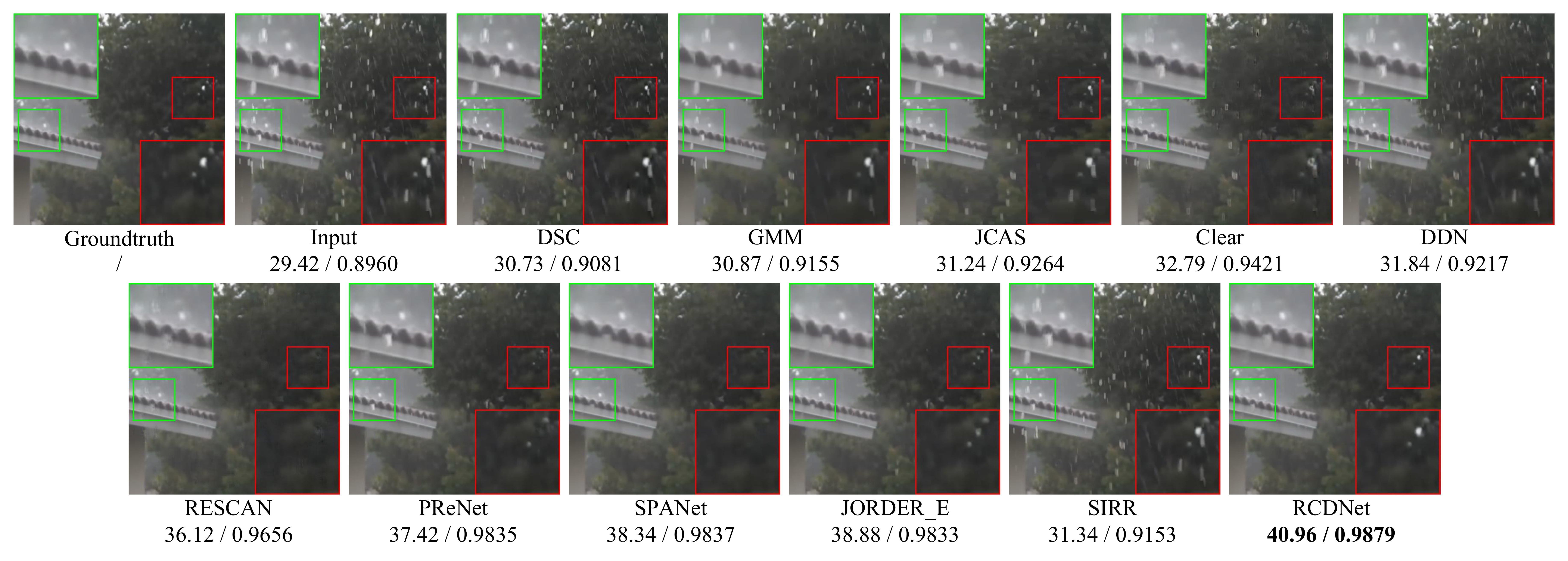}
  \end{center}
  \vspace{-5mm}
     \caption{Rain removal performance comparisons on a rainy image from SPA-Data. The images are better to be zoomed in on screen.}
  \label{figwang}
    \vspace{-3mm}
\end{figure*}
\begin{figure*}[t]
  \begin{center}
  \vspace{-1mm}
     \includegraphics[width=0.82\linewidth]{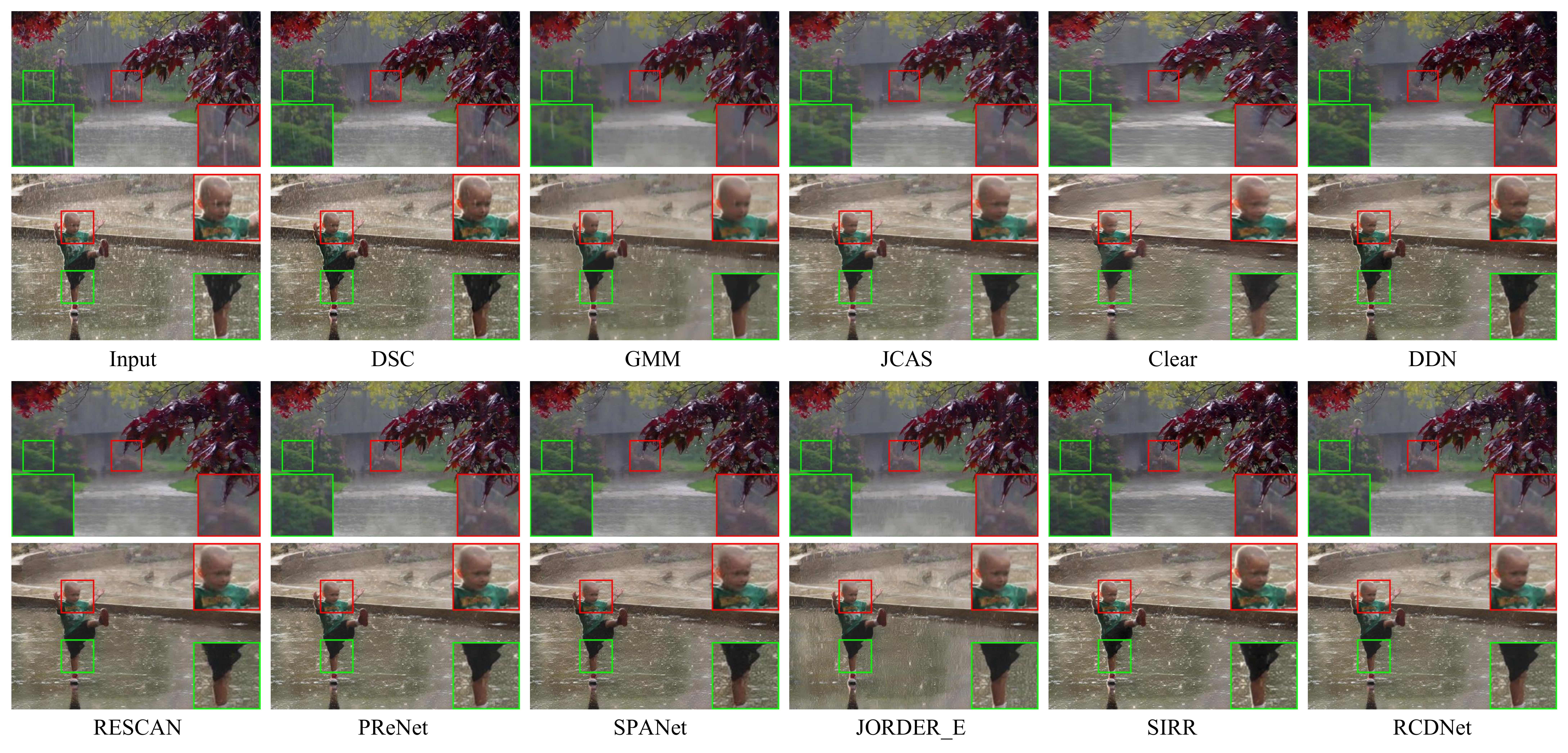}
  \end{center}
  \vspace{-5mm}
     \caption{Derained results for two samples with various rain patterns from Internet-Data. The images are better to be zoomed in on screen.}
  \label{figreal}
    \vspace{-6mm}
\end{figure*}
\subsection{Experiments on synthetic data}
\textbf{Comparison methods and datasets.}
We then compare our network with the current SOTA single image derainers, including model-based DSC~\cite{Yu2015Removing}, GMM~\cite{Li2016Rain}, and JCAS~\cite{Gu2017Joint}; DL-based Clear~\cite{Fu2017Clearing}, DDN~\cite{Fu2017Removing}, RESCAN~\cite{li2018recurrent}, PReNet~\cite{ren2019progressive}, SPANet~\cite{wang2019spatial}, JORDER\_E~\cite{Yang2019Joint}, and SIRR~\cite{wei2019semi}\footnote{The code/project links for these comparison methods are listed in supplementary material.}, based on four benchmark datasets, including Rain100L, Rain100H~\cite{Yang2019Joint}, Rain1400~\cite{Fu2017Removing}, and Rain12~\cite{Li2016Rain}.

Fig.~\ref{figrainl} illustrates the deraining performance of all competing methods on a rainy image from Rain100L. As shown, the deraining result of RCDNet is better than that of other methods in sufficiently removing the rain streaks and finely recovering the image textures. Moreover, the rain layer extracted by RCDNet contains fewer unexpected background details as compared with other competing methods. Our RCNet thus achieves the best PSNR and SSIM.

Table~\ref{tabsyn1} reports the quantitative results of all competing methods. It is seen that our RCDNet attains best deraining performance among all methods on each dataset. This substantiates the flexibility and generality of our method, in diverse rain types contained in these datasets.
\subsection{Experiments on real data}
We then analyze the performance of all methods on two real datasets from~\cite{wang2019spatial}: the first one (called SPA-Data) contains 638492 rainy/clean image pairs for training and 1000 testing ones, and the second one (called Internet-Data) includes 147 rainy images without groundtruth.

Table~\ref{tabwang1} and Fig.~\ref{figwang} compare the derained results on SPA-Data of all competing methods visually and quantitatively. It is easy to see that even for such complex rain patterns, the proposed RCDNet still achieves an evident superior performance than other methods. Especially, similar to its superiority in synthetic experiments, it is also observed that our method better removes the rain streaks and recovers image details than other competing ones.
\begin{table}[t]
\vspace{2mm}
\centering
\caption{PSNR and SSIM comparisons on SPA-Data~\cite{wang2019spatial}.}
\footnotesize
\setlength{\tabcolsep}{2pt}
\begin{tabular}{p{1cm}<{\centering}|p{1cm}<{\centering}p{1cm}<{\centering}p{1cm}<{\centering}p{1.2cm}<{\centering}p{1cm}<{\centering}p{1cm}<{\centering}}
\hline
 Methods &Input  &DSC  &GMM &JCAS &Clear &DDN\\
\hline
 PSNR &34.15     &34.95   &34.30   &34.95  &34.39   &36.16  \\
\hline
 SSIM &0.9269    &0.9416  &0.9428 &0.9453   &0.9509   &0.9463 \\
\hline
 Methods & RESCAN &PReNet  &SPANet  &JORDER\_E  &SIRR  &RCDNet\\
\hline
 PSNR &38.11    &40.16     &40.24    &\textit{\textbf{40.78}}  & 35.31  &{\textbf{41.47}}  \\
\hline
 SSIM &0.9707   &\emph{\textbf{0.9816}}  &0.9811  &0.9811   &0.9411  &\textbf{{0.9834}} \\
\hline
\end{tabular}
\label{tabwang1}
\vspace{-6mm}
\end{table}

Further, we select two real hard samples with various rain densities to evaluate the generalization ability of all competing methods.  From Fig.~\ref{figreal}, we can find that traditional model-based methods tend to leave obvious rain streaks. Although DL-based comparison methods remove apparent rain streaks, they still leave distinct rain marks or blur some image textures. Comparatively, our RCDNet better preserves background details as well as removes more rain streaks. This shows its good generalization capability to unseen complex rain types.
\section{Conclusion}
\vspace{-2mm}
In this paper, we have explored the intrinsic prior structure of rain streaks that can be explicitly expressed as convolutional dictionary learning model, and proposed a novel interpretable network architecture for single image deraining. Each module in the network can one-to-one correspond to the implementation operators of the algorithm designed for solving the model, and thus the network is almost ``white-box" with easily visualized interpretation for all its module elements. Comprehensive experiments implemented on synthetic and real rainy images validate that such interpretability brings a good effect of the proposed network, and especially facilitates the analysis for how it happens in the network and why it works in testing prediction process. The extracted elements through the end-to-end learning by the network, like the rain kernels, are also potentially useful for the related tasks on rainy images.

\noindent
\textbf{Acknowledgment.} This research was supported by
the China NSFC projects under contract 11690011, 61721002, U1811461 and MoE-CMCC ``Artifical Intelligence" Project with No. MCM20190701

\newpage
{\small
\bibliographystyle{ieee_fullname}
\bibliography{egbib}
}
\end{document}